\documentclass[lettersize,journal]{IEEEtran}
\usepackage{amsmath,amsfonts}
\usepackage{algorithmic}
\usepackage{algorithm}
\usepackage{array}
\usepackage[caption=false,font=normalsize,labelfont=sf,textfont=sf]{subfig}
\usepackage{textcomp}
\usepackage{stfloats}
\usepackage{url}
\usepackage{verbatim}
\usepackage{graphicx}
\usepackage{cite}
\begin{document}

\title{Spectro-Temporal Modulation Representation Framework for  Human-Imitated Speech Detection}

\author{Khalid Zaman and Masashi Unoki
        % <-this % stops a space
\thanks{Khalid Zaman and Masashi Unoki are with the Graduate School of Advanced Science and Technology, Japan Advanced Institute of Science and Technology, Nomi, Ishikawa 923-1292, Japan (e-mail: zaman.khalid@jaist.ac.jp; unoki@jaist.ac.jp).}% <-this % stops a space
}



\maketitle

\begin{abstract}
Human-imitated speech poses a greater challenge than AI-generated speech for both human listeners and automatic detection systems. Unlike AI-generated speech, which often contains artifacts, over-smoothed spectra, or robotic cues, imitated speech is produced naturally by humans, thereby preserving a higher degree of naturalness that makes imitation-based speech forgery significantly more challenging to detect using conventional acoustic or cepstral features. To overcome this challenge, this study proposes an auditory perception-based Spectro-Temporal Modulation (STM) representation framework for human-imitated speech detection. The STM representations are derived from two cochlear filterbank models: the Gammatone Filterbank (GTFB), which simulates frequency selectivity and can be regarded as a first approximation of cochlear filtering, and the Gammachirp Filterbank (GCFB), which further models both frequency selectivity and level-dependent asymmetry. These STM representations jointly capture temporal and spectral fluctuations in speech signals, corresponding to changes over time in the spectrogram and variations along the frequency axis related to human auditory perception. We also introduce a Segmental-STM representation to analyze short-term modulation patterns across overlapping time windows, enabling high-resolution modeling of temporal speech variations. Experimental results show that STM representations are effective for human-imitated speech detection, achieving accuracy levels close to those of human listeners. In addition, Segmental-STM representations are more effective, surpassing human perceptual performance. The findings demonstrate that perceptually inspired spectro-temporal modeling is promising for detecting imitation-based speech attacks and improving voice authentication robustness.
\end{abstract}


\section{Introduction}
\IEEEPARstart{S}{peech} is widely used in modern technologies such as voice assistants, biometric systems, and online communication platforms. While these technologies have improved convenience and accessibility for users, they have also introduced new security concerns. In particular, speech signals have become increasingly vulnerable to manipulation and forgery, creating significant challenges for speaker verification, authentication, and privacy protection~\cite{marras2023dictionary,wubet2025speaker,khan2023securing,ballesteros2021deep4snet,unoki2024deepfake,zaman2024hybrid,kanwal2024fake}. Among these emerging threats, AI-generated synthetic speech has attracted significant attention, as advanced techniques such as text-to-speech (TTS), voice conversion (VC), and deep learning–based synthesis can now produce highly realistic and natural-sounding speech. These synthesis approaches have rapidly advanced with the introduction of neural architectures such as Tacotron~\cite{wang2017tacotron, shen2018natural}, WaveNet~\cite{van2016wavenet}, and GAN-based~\cite{yamamoto2020parallel,kumar2019melgan} and diffusion-based vocoders~\cite{kong2020diffwave}, which can generate highly natural and expressive speech by reproducing speaker-specific traits and emotional tones. Trained on large-scale speech corpora, these data-driven models learn to capture prosodic patterns, pitch contours, and vocal timbre with remarkable fidelity.

To systematically evaluate the security implications of these technological advances, several benchmark corpora and detection challenges have been established, including ASVSpoof \cite{kinnunen2018automatic,kinnunen20182nd,kinnunen2017asvspoof,wang2020asvspoof,liu2023asvspoof,wang2024asvspoof}, audio deepfake detection (ADD) \cite{yi2022add}, \cite{yi2023add}, Real or Fake \cite{reimao2019dataset}, Chinese Fake Audio Detection (CFAD) \cite{ma2024cfad}, Wavfake \cite{zaidi2021touch}, and In-the-Wild (ITW) Wavfake \cite{muller2022does}, along with related spoofing studies \cite{yi2021half, frank2021wavefake, kominek2004cmu, bird2023real, zhang2022partialspoof, xie2025codecfake, zhang2021fmfcc, lorenzo2018voice}. These benchmarks have played a crucial role in identifying vulnerabilities in automatic speaker verification systems and in facilitating the development and comparison of anti-spoofing techniques under diverse attack scenarios and acoustic conditions. Despite the success of these datasets and challenges, most AI-generated speech still exhibits detectable artifacts. Such signals often sound relatively uniform or slightly robotic, with imperfections that arise from over-smoothed spectral envelopes, limited fine-grained prosodic variation, and temporal or phase distortions introduced by statistical averaging or model regularization. Consequently, although synthetic speech may sound natural at a coarse level, it typically lacks the subtle micro-level articulatory and modulation cues inherent to genuine human speech, making it comparatively easier for current detection systems to identify.

In contrast, human-imitated speech presents a substantially more challenging detection problem. Human-imitated speech, produced by humans intentionally mimicking another speaker \cite{almutairi2022review}, closely resembles genuine speech in its naturalness and expressiveness. It preserves pitch dynamics, rhythm, articulation patterns, and vocal timbre with high fidelity, often making it difficult for both human listeners and automatic systems to distinguish from genuine speech \cite{unoki2024deepfake}, \cite{almutairi2022review}. This challenge is further exacerbated by the scarcity of datasets specifically designed for imitation-based detection \cite{unoki2024deepfake}. Unlike synthetic or voice-converted speech, genuine human imitation datasets are limited in size, lack standardized recording conditions, and are underrepresented in current benchmarks. As a result, existing spoofing detection systems, which are largely optimized for algorithmically generated speech, tend to perform poorly when confronted with naturally produced human-imitated speech. This performance gap highlights a critical limitation of current approaches and motivates the development of detection methods that better reflect perceptually relevant characteristics of speech signals.

While several studies have attempted to address imitation-related detection, most prior work has focused on vocal or synthetic imitation rather than genuine human-imitated speech. For example, the vocal imitation dataset \cite{kim2018vocal} and related signal-processing-based approaches \cite{rodriguez2020machine} primarily investigate artificial or sound-level imitation. Other studies examine non-speech imitation, such as the Arabic cantillation corpus \cite{lataifeh2020arabic}, or musical imitation, as in the vocal percussion corpus \cite{mehrabi2019vocal}. Zetterholm \cite{zetterholm2007detection} analyzed voice imitation using a small set of imitations of Swedish public figures, combining perceptual listening tests with acoustic analyses of features such as fundamental frequency, formants, articulation rate, and spectral characteristics of fricatives. However, the limited dataset size and lack of statistical modeling constrained the generalizability of the findings. Similarly, Hautamäki et al.~\cite{hautamaki2015automatic} evaluated a Finnish mimicry corpus using both automatic speaker verification systems and human listeners, while Zaman et al.~\cite{zaman2025ability} examined human-imitated speech through subjective listening experiments and timbral feature analysis. However, both studies relied on limited acoustic representations, resulting in modest detection performance. Collectively, the existing methods often depend on descriptive analyses or low-level spectral features, which are insufficient to capture the subtle perceptual cues that distinguish genuine speech from high-quality human imitation. This observation underscores the need for more advanced and perceptually grounded acoustic representations.

Temporal and spectral modulations within the modulation domain provide a representation of acoustic variations across time and frequency, as proposed by Flinker et al.~\cite{flinker2019spectrotemporal}. Accordingly, recent studies in speech and audio analysis have shown that modeling spectro-temporal modulations (STM) provides a richer and more perceptually relevant representation of sound than conventional spectral or cepstral features. Wu et al.~\cite{wu2013synthetic} introduced magnitude and phase modulation features for detecting synthetic speech, demonstrating that modulation-domain information improves spoofing detection performance. STM representations have also been applied to deepfake speech detection; for example, Cheng et al.~\cite{cheng2023analysis} reported that STM features better simulate aspects of human auditory perception and enhance discrimination between genuine and fake speech by capturing subtle articulatory and prosodic cues. In addition, Li et al.~\cite{li2025machine} proposed an STM analysis framework with machine-specific filterbanks for anomaly detection. In this method, they first quantified machine-specific frequency importance using the Fisher ratio (F-ratio) and designed non-uniform filterbanks (NUFBs) to extract the Log Non-Uniform Spectrum (LNS), emphasizing discriminative frequency regions. Spectral and temporal modulation representations derived from the LNS features were then fed into an autoencoder-based detector to improve anomaly detection performance across different SNR conditions. These findings suggest that STM analysis offers a perceptually grounded framework aligned with auditory processing mechanisms in the human brain, making it particularly well suited for addressing the challenges of human-imitated speech detection.

In the present study, we aim to achieve more robust detection of human-imitated speech by utilizing acoustic features related to auditory perception and informed by the subjective listening experiments reported in \cite{zaman2025ability}. Specifically, we employ STM representations derived from Gammatone Filterbank (GTFB) and Gammachirp Filterbank (GCFB) outputs to model cochlear filtering and auditory modulation patterns. Here, GTFB can be regarded as a first approximation of cochlear filtering that simulates frequency selectivity, while GCFB further models both frequency selectivity and level-dependent asymmetry. We also introduce Segmental-STM features that capture short-term modulation patterns across speech segments, enabling segment-level temporal and spectral analysis. Our evaluations of these auditory-inspired representations utilizing multiple machine-learning classifiers demonstrate an improved detection performance compared to conventional acoustic features.

The primary novelty of this work lies in the application of auditory-inspired STM and Segmental-STM features derived from gammatone and gammachirp filterbank-based auditory representations to the detection of naturally produced human-imitated speech, a problem that remains largely unexplored. These representations capture perceptually relevant spectro-temporal patterns and short-term modulation variations, enabling an interpretable and effective approach for distinguishing genuine speech from human imitation.

Section~\ref{Proposed-methods} of this paper describes the computation of the proposed STM representations and the associated machine-learning modules. Section~\ref{Experiments} presents the datasets, experimental setup, and evaluation metrics. The results and discussion are provided in Section~\ref{sec:results_discussion}, followed by a general discussion in Section~\ref{General Discussion}. Finally, concluding remarks and future directions are given in Section~\ref{Conclusion}.

\begin{figure*}[!t]
\centering
\includegraphics[width=0.96\linewidth]{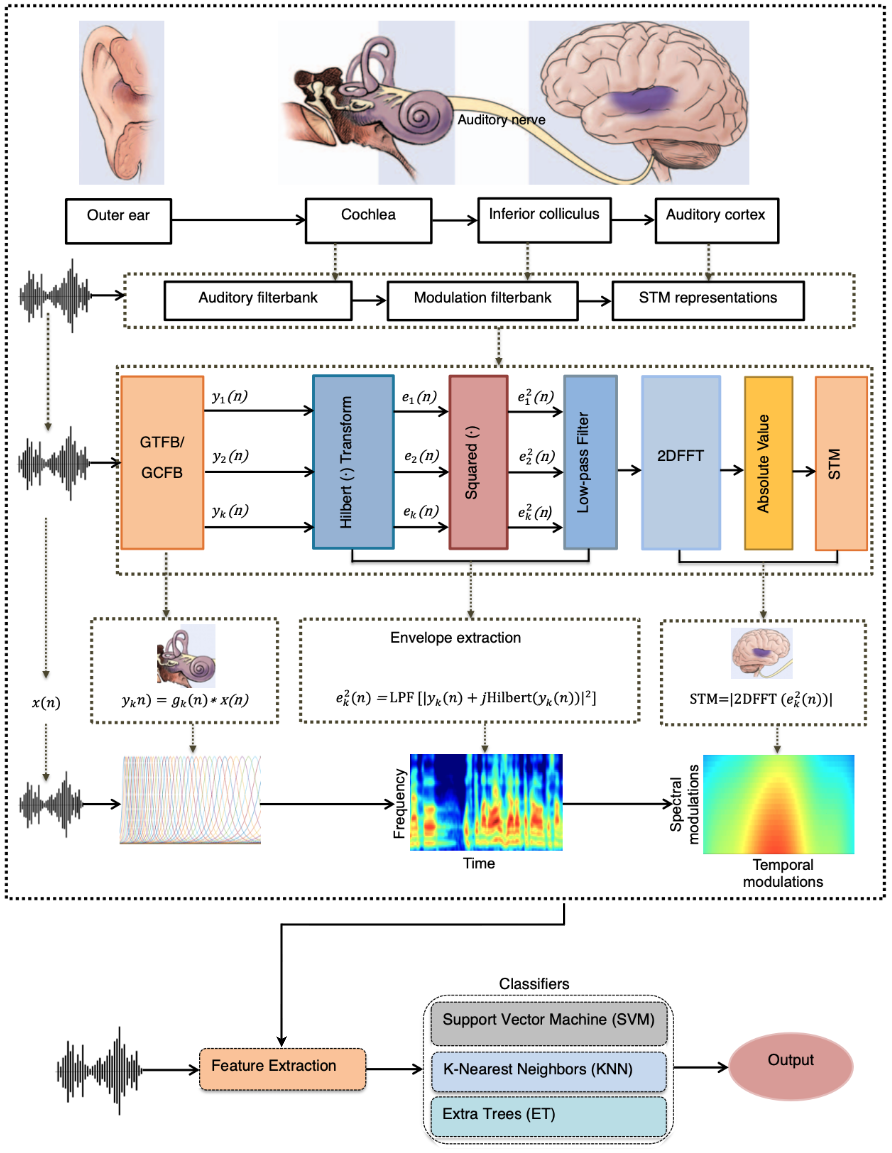}
\caption{Block diagram of STM representation framework illustrating the correspondence between human auditory processing and its computational modeling.}
\label{fig1}
\end{figure*}

\section{Proposed Methods}
\label{Proposed-methods}
This study extends our previous work~\cite{zaman2025ability} by introducing auditory-inspired STM representations to capture spectral and temporal variations in speech signals. Motivated by human auditory perception, these representations emphasize modulation patterns across time and frequency that play an important role in distinguishing genuine speech from human-imitated speech and are not fully captured by conventional spectral features. 

The conceptual correspondence between the human auditory processing mechanism \cite{bance2007hearing,li2025machine} and its computational modeling is illustrated in Fig.~\ref{fig1}. In the auditory system, incoming speech is captured by the outer ear and decomposed into frequency-selective responses in the cochlea. The resulting signals are transmitted through the auditory nerve to midbrain structures such as the inferior colliculus and further processed in the auditory cortex, where sensitivity to spectro-temporal modulations is well established \cite{li2025machine, chi2012robust}.

Inspired by this hierarchical processing pathway, the proposed framework models cochlear frequency analysis using an auditory filterbank based on gammatone and gammachirp filters that simulate the frequency selectivity of the cochlea. This process produces subband signals corresponding to different frequency channels. For each subband, the envelope is extracted utilizing the Hilbert transform followed by low-pass filtering, capturing slow temporal amplitude variations. These envelope signals collectively form an auditory spectrogram that represents the time–frequency energy distribution of the input signal.

To model cortical processing, the auditory spectrogram is further analyzed using a modulation filterbank implemented via a two-dimensional discrete Fourier transform (2D-DFT). This enables joint analysis of spectral modulation (SM) and temporal modulation (TM) components. The resulting spectro-temporal modulation captures the distribution of spectro-temporal modulation energy, providing a dynamic representation of speech that reflects underlying modulation patterns.

% The auditory spectrogram and STM representation spectrogram are adopted from \cite{flinker2019spectrotemporal}, as shown in Fig.~\ref{fig1}.

We also introduce segmental-STM representations, which compute STM representations over short, overlapping time windows to enable segment-level analysis of dynamic articulatory variations. This segmental modeling better captures short-term modulation patterns that may differ between genuine and human-imitated speech.

To evaluate the effectiveness of the proposed representations, we utilize three machine-learning classifiers to classify genuine and human-imitated speech samples: Support Vector Machine (SVM), k-Nearest Neighbors (KNN), and Extra Trees (ET). The following subsections describe the STM computation and classifier configurations in detail.

\subsection{Spectro-Temporal Modulation (STM) Representation}
In this study, two auditory filterbanks were employed to derive the STM representations of speech: the GTFB ~\cite{patterson1988efficient} and the GCFB \cite{irino1997gammachirp} \cite{irino1999analysis}. Both front-ends simulate the auditory frequency selectivity of the human cochlea and were followed by a consistent STM extraction pipeline comprising envelope computation, modulation-domain transformation, and segmental analysis. The overall processing procedure is illustrated in Fig.~\ref{fig1}.

\subsubsection{STM Derived from the GTFB}
Spectro-temporal modulation (STM) representations are computed through a multi-stage process comprising auditory filtering, envelope demodulation, and modulation spectral analysis. First, a gammatone auditory filterbank is applied to decompose the discrete-time speech signal $x[n]$ into multiple frequency subbands that approximate the critical-band structure of the human cochlea. The center frequencies and bandwidths follow the equivalent rectangular bandwidth (ERB) scale:
\begin{equation}
\mathrm{ERB}(f_k) = 24.7 \left(4.37 f_k + 1 \right),
\end{equation}
where $f_k$ (in kHz) denotes the center frequency of the $k$-th filter. In this work, $64$ channels were used to span the frequency range from $60$ Hz to $7.6$ kHz, providing an acoustic representation analogous to the auditory periphery.

The impulse response of the $k$-th gammatone filter is given by
\begin{equation}
\begin{aligned}
g_k[n] &= \mathrm{A} (nT_s)^{p-1} \exp\left(-2\pi b_f \,\mathrm{ERB}(f_k)\, nT_s \right) \\
&\quad \times \cos\left(2\pi f_k nT_s\right), \quad n \ge 0,
\end{aligned}
\end{equation}
where $A$ is a gain constant, $p$ is the filter order ($p = 4$ in this study), $b_f$ is a bandwidth scaling factor, $F_s$ is the sampling frequency, and $T_s = 1/F_s$ is the sampling period.

The output of the $k$-th auditory channel is obtained through discrete convolution between the input speech signal $x[n]$ and the impulse response of the corresponding filter:
\begin{equation}
y_k[n] = (x * g_k)[n],
\end{equation}
where $*$ denotes convolution and $g_k[n]$ represents the impulse response of the $k$-th filter.

To estimate the envelope, the power envelope of each subband signal is computed using the Hilbert transform. The squared absolute value of $y_k[n] + j\,\mathrm{Hilbert}\{y_k[n]\}$ is then obtained, followed by low-pass filtering to preserve low-frequency modulation components.

\begin{equation}
e_k^2[n] = \mathrm{LPF} \left( \left| y_k[n] + j\,\mathrm{Hilbert}\left( y_k[n] \right) \right|^2 \right),
\end{equation}

where $\mathrm{Hilbert}\{\cdot\}$ denotes the Hilbert transform and $\mathrm{LPF}$ is a low-pass filter with a cutoff frequency of $64$ Hz.

To obtain a joint representation of spectral and temporal modulations, a two-dimensional fast Fourier transform (2D-FFT) is applied across the auditory channel index $k$ and time index $n$:
\begin{equation}
\mathrm{STM}_{\mathrm{GTFB}} = \mathrm{2DFFT} \left( e_k^2[n] \right),
\end{equation}
where $\mathrm{2DFFT}$ denotes the two-dimensional fast Fourier transform. The transform produces a complex-valued matrix, and the STM representation is obtained by taking its absolute value.

The resulting $\mathrm{STM}_{\mathrm{GTFB}}$ characterizes modulation energy across spectral and temporal dimensions, capturing perceptually relevant speech patterns such as formant transitions, phoneme dynamics, and voicing structures.

\subsubsection{STM Derived from the GCFB}
While the GTFB provides symmetric auditory filters, it does not fully account for the asymmetric and level-dependent tuning characteristics of the cochlea. To incorporate more realistic auditory filtering behavior, the gammatone front-end is replaced with a GCFB. The GCFB introduces a frequency-dependent chirp term into the filter response, enabling asymmetric auditory tuning and nonlinear compression.

The impulse response of the $c$-th gammachirp filter in discrete time is given by
\begin{equation}
\begin{split}
g_k[n] = \mathrm{A} (nT_s)^{p-1} 
\exp\!\left(-2\pi b_f\,\mathrm{ERB}(f_c)\,nT_s\right) \\
\cos\!\left(2\pi f_c nT_s + c\,\ln(nT_s)\right), \quad n \ge 0,
\end{split}
\end{equation}

\noindent where $A$ denotes the gain constant, $p$ is the filter order (set to 4), $b_f$ determines the effective bandwidth, $f_c$ denotes the center frequency of the $c$-th filter, and $c$ is the chirp coefficient controlling the degree of frequency glide and asymmetry. The sampling frequency is denoted by $F_s$, and $T_s = 1/F_s$ represents the sampling period. This formulation allows the GCFB to more accurately simulate the upward and downward spectral spread observed in cochlear mechanics, providing improved modeling of real auditory filter responses.

Following auditory decomposition, the subband signal corresponding to the $c$-th gammachirp channel is obtained by discrete convolution, as
\begin{equation}
y_k[n] = (x * g_k)[n], 
\end{equation}
where $*$ denotes discrete-time convolution.

After auditory filtering, the same envelope extraction and spectro-temporal modulation analysis procedures described for the GTFB case are applied. Specifically, the power envelope is computed using the Hilbert transform by calculating the squared absolute value of the complex representation, followed by low-pass filtering to preserve low-frequency modulation components. A two-dimensional fast Fourier transform (2D-FFT) is then applied across time and frequency channel dimensions to obtain the STM representation.

\begin{equation}
\mathrm{STM_{GCFB}} = \left| \mathrm{2DFFT}\left(e_k^2[n]\right) \right|
\end{equation}

The resulting $\mathrm{STM}_{\mathrm{GCFB}}$  captures the spectro-temporal modulation representation within an auditory framework that accounts for nonlinear loudness growth, frequency glide, and masking asymmetry, thereby providing a perceptually richer acoustic representation.

\subsubsection{Segmental-STM Representation}

\begin{figure}[t]
\begin{center}
\includegraphics[width=1\linewidth]{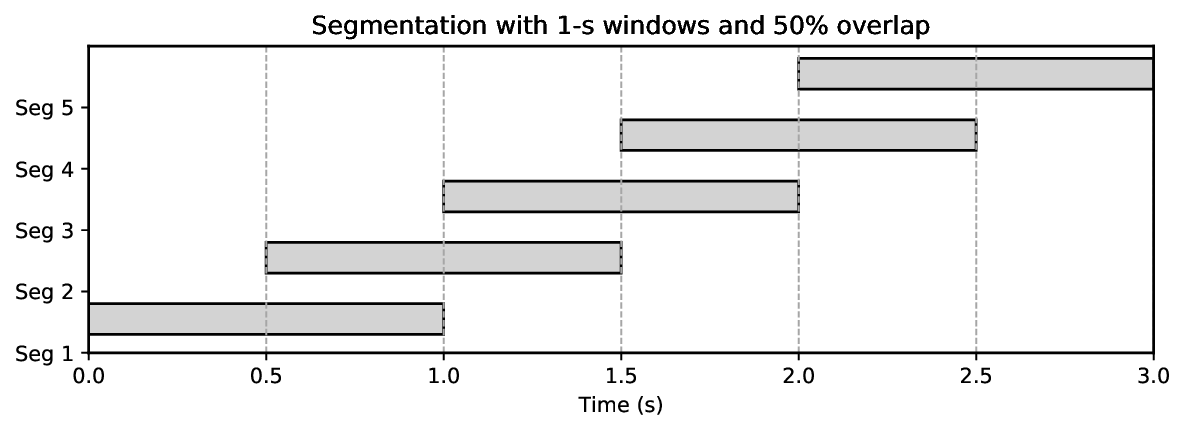}
\end{center}
\caption{Segmentation scheme with 1-s windows and 50\% overlap.}
\label{fig2}
\end{figure}

Speech acoustics are inherently non-stationary, as they exhibit time-varying amplitude and spectral modulation patterns. To capture these temporal dynamics, we introduced a \textit{segmental-STM} representation and applied it to both GTFB- and GCFB-based systems, as shown in Fig.~\ref{fig2}. The modulation envelopes were resampled to a rate of 160~Hz to constrain the modulation bandwidth and avoid aliasing. The resampled envelopes were then divided into overlapping temporal segments of 1-s duration with a 50\% overlap. Each segment was processed independently using the same auditory modulation analysis pipeline as described earlier in this study.

For the $i$-th temporal segment, the segment-wise spectro-temporal modulation spectrum is computed as

\begin{equation}
\mathrm{STM}_{\mathrm{seg}}(i) =
\left|
\mathrm{2DFFT}\!\left(e_k^2[n]\right)
\right|,
\quad n \in \text{segment } i,
\end{equation}
where $e_k^2[n]$ denotes the discrete-time power envelope of the $k$-th auditory channel within the $i$-th temporal segment, and $k = 1,2,\ldots,K$ denotes the auditory channel index.

All segmental-STM representations are concatenated along the temporal axis to form a three-dimensional tensor:

\begin{equation}
\mathrm{STM}_{\mathrm{seg}} \in \mathbb{R}^{K \times M \times S},
\end{equation}
where $K$ represents the number of auditory channels (filters), $M$ denotes the number of spectro-temporal modulation frequency bins obtained after the two-dimensional FFT operation, and $S$ is the total number of temporal segments.

This segment-level representation captures spectro-temporal modulation patterns over short time intervals, providing enhanced temporal resolution and improved discrimination of acoustic variations in speech signals. Consequently, the segmental-STM serves as a more discriminative acoustic descriptor that complements the utterance-level STM representation in distinguishing genuine and imitated speech. When these representations are extracted from the GTFB and GCFB, the corresponding features are referred to as $\mathrm{STM}_{\mathrm{seg}}$ (GTFB) and $\mathrm{STM}_{\mathrm{seg}}$ (GCFB), respectively.

\subsection{Machine Learning Models}
\label{sec:ml-models}
In this study, we utilized three machine learning models to evaluate the proposed methods.

\paragraph{Support Vector Machine (SVM).} Support Vector Machine (SVM) is a supervised learning algorithm widely used for binary classification tasks. Its primary objective is to construct an optimal separating hyperplane that maximizes the margin between samples of two classes. When the training data are linearly separable, this hyperplane can be defined directly in the original feature space.

However, real-world data often exhibit complex, non-linear distributions that cannot be effectively separated using a linear decision boundary. To address this limitation, SVM can be extended through the use of kernel functions, which implicitly map the input features into a higher-dimensional feature space where linear separation becomes feasible.

In this work, we employed a non-linear radial basis function (RBF) kernel to model complex decision boundaries between genuine and imitated speech samples. The RBF kernel enables the classifier to implicitly project the input features into a high-dimensional space, allowing for more effective separation of the two classes when the data are not linearly separable in the original input space.

Given a set of training samples $\{(\mathbf{x}_i, y_i)\}_{i=1}^{N}$ with class labels $y_i \in \{-1,+1\}$, the SVM decision function in its dual form is expressed as
\begin{equation}
f(\mathbf{x}) = \mathrm{sign}\!\left(\sum_{i=1}^{N} \alpha_i y_i K(\mathbf{x},\mathbf{x}_i) + b_0\right),
\end{equation}
where $\alpha_i$ denotes the Lagrange multipliers associated with the support vectors, $b_0$ is the bias term, and $K(\cdot,\cdot)$ represents the kernel function.

Among various kernel choices, the radial basis function (RBF) kernel is defined as
\begin{equation}
K(\mathbf{x},\mathbf{x}_i) = \exp\!\left(-\gamma \lVert \mathbf{x} - \mathbf{x}_i \rVert_2^2\right),
\end{equation}
where $\gamma > 0$ controls the width of the Gaussian kernel and determines the locality and smoothness of the resulting decision boundary.

\paragraph{k-Nearest Neighbors (KNN).} The KNN classifier assigns a test sample to the most common class among its closest neighbors in the feature space. This distance-based approach is simple yet effective in capturing local structures and similarity relationships between samples without assuming any explicit distribution of the data.

Let $\mathcal{N}_k(\mathbf{x})$ denote the set of indices of the $k$ nearest neighbors of a test sample $\mathbf{x}$. The predicted class label is given by
\begin{equation}
\hat{y} = \arg\max_{\ell \in \mathcal{C}} \sum_{i \in \mathcal{N}_k(\mathbf{x})} \mathbb{I}(y_i = \ell),
\end{equation}

where $\mathcal{C}$ is the set of class labels, $\ell \in \mathcal{C}$ denotes a class label, and $\mathbb{I}(\cdot)$ is the indicator function.

The distance between samples is computed using the Euclidean metric:
\begin{equation}
d(\mathbf{x},\mathbf{x}_i) = \lVert \mathbf{x} - \mathbf{x}_i \rVert_2.
\end{equation}

\paragraph{Extra Trees (ET).} The Extra Trees classifier, an ensemble-based learning algorithm, was used to evaluate the contribution of randomization and feature sub-sampling. It constructs multiple uncorrelated decision trees utilizing random splits and aggregates their predictions, thereby reducing overfitting and improving robustness across diverse acoustic feature sets.

Given an ensemble of $T$ trees, the final prediction is obtained by averaging class posterior probabilities, as
\begin{equation}
\hat{P}(y=\ell \mid \mathbf{x}) = \frac{1}{T} \sum_{t=1}^{T} \hat{P}_t(y=\ell \mid \mathbf{x}),
\end{equation}
and assigning the class label
\begin{equation}
\hat{y} = \arg\max_{\ell \in \mathcal{C}} \hat{P}(y=\ell \mid \mathbf{x}),
\end{equation}
where $\hat{P}_t(y=\ell \mid \mathbf{x})$ denotes the posterior probability of class $\ell$ estimated by the $t$-th tree.

At each node $u$, a random subset of features $\mathcal{F}_u$ is selected, and for each feature $m \in \mathcal{F}_u$, a split threshold is sampled uniformly as
\begin{equation}
\theta_m \sim \mathcal{U}\!\left(\min_{i \in S_u} x_{im}, \; \max_{i \in S_u} x_{im}\right),
\end{equation}
where $S_u$ denotes the set of samples reaching node $u$.

Among the randomly generated candidate splits, the optimal split is selected by maximizing the impurity reduction:
\begin{equation}
\Delta I = I(S_u) - \frac{|S_L|}{|S_u|} I(S_L) - \frac{|S_R|}{|S_u|} I(S_R),
\end{equation}
where $S_L$ and $S_R$ represent the left and right child nodes, respectively. The impurity is measured using the Gini index:
\begin{equation}
I(S) = 1 - \sum_{\ell \in \mathcal{C}} p_{\ell}(S)^2,
\end{equation}
with
\begin{equation}
p_{\ell}(S) = \frac{1}{|S|} \sum_{i \in S} \mathbb{I}(y_i = \ell).
\end{equation}

\section{Experiments}
\label{Experiments}
\subsection{Dataset and Evaluation Metrics}
The experiments in this study utilized the Human-Imitated Speech dataset introduced in our previous work~\cite{zaman2025ability}. The dataset includes genuine and imitated speech produced by professional artists across multiple languages for ten target speakers. A total of 100 speech samples were used, equally divided between genuine and imitated speech. Audio segments were manually extracted from publicly available online videos, with only the less noisy portions retained to ensure clarity. Despite challenges such as background noise and limited availability of high-quality imitation samples, the final dataset provides diverse and realistic speech conditions. Model performance was evaluated using accuracy and the confusion matrix.

\subsection{Training Details}
A total of 40 speech samples were used for training and 100 samples for testing. Two types of STM representations were extracted from each utterance. The first type, referred to as the STM, was obtained by applying a 2D Fast Fourier Transform (2D-FFT) to the full-length resampled auditory envelope, producing a two-dimensional modulation map with a dimension of $64 \times 480$. This representation captures the overall spectral and temporal modulation patterns of the entire utterance. 

The second type, the Segmental-STM, was generated by dividing the modulation envelope into overlapping temporal segments of 1-s duration with a 50\% overlap and performing 2D-FFT on each segment. The resulting STM maps were concatenated into a three-dimensional tensor of size $64 \times 160 \times S$, where $S$ denotes the number of segments. 

This approach captures temporal variations in spectro-temporal modulation patterns, thus complementing the global STM representation.

All STM features were computed utilizing auditory filterbank front-ends based on the GTFB and GCFB models, followed by envelope extraction, resampling to 160~Hz, and magnitude computation after 2D-FFT. These features were then flattened and standardized before being used as input to the SVM, KNN, and Extra Trees classifiers for imitated speech detection.

\section{Results and Discussion}
\label{sec:results_discussion}
\subsection{STM Evaluation}
Table~\ref{tab:stm_global} presents the results for STM-based acoustic features derived from the GTFB and GCFB filterbanks. Across all classifiers, the STM(GCFB) features consistently outperformed STM(GTFB), highlighting the effectiveness of the gammachirp representation, which more closely aligns with human auditory filter characteristics. The STM(GCFB) feature representation with the KNN classifier achieved the highest overall performance, reaching 69\% accuracy. This result indicates that STM(GCFB) features effectively capture spectro–temporal patterns that distinguish genuine and imitated speech, while the distance-based modeling of KNN further supports discrimination, achieving higher performance than margin-based or tree-based classifiers. These findings demonstrate that modulation cues extracted over the entire utterance encode perceptually meaningful information, even without segmental decomposition.

\begin{table}[htbp]
  \centering
 \caption{Performance of Global STM (GTFB and GCFB) Features using Different Classifiers.}
  \label{tab:stm_global}
  \small
  \setlength{\tabcolsep}{6pt}
  \renewcommand{\arraystretch}{1.3}
  \begin{tabular}{l c c}
    \hline
 \textbf{Feature Type} & \textbf{Classifier} & \textbf{Accuracy (\%)} \\ 
    \hline
    STM(GTFB) & SVM & 61.0 \\
    STM(GCFB) & SVM & 62.0 \\
    STM(GTFB) & KNN & 68.0 \\
    STM(GCFB) & KNN & \textbf{69.0} \\
    STM(GTFB) & Extra Trees & 63.0 \\
    STM(GCFB) & Extra Trees & 62.0 \\
    \hline
  \end{tabular}
\end{table}

\subsection{Segmental-STM Evaluation}
When the modulation analysis was applied segmentally, as shown in Table~\ref{tab:stm_segmental}, the performance improved further. By computing modulation spectra over short overlapping windows, segmental-STM captures short-term and time-varying modulation patterns that are often averaged out when features are extracted from the entire utterance. As shown in the table, the $\mathrm{STM}_{\mathrm{seg}}$(GCFB) feature representation with the Extra Trees classifier achieved the highest accuracy of 71\%, indicating that incorporating short-term spectro–temporal information improves the representation of dynamic acoustic variations. Although SVM and KNN produced moderate results, Extra Trees showed better performance when segment-level STM features were used, suggesting that it benefits more from segmentation.

\begin{table}[htbp]
  \centering
 \caption{Performance of Segmental-STM (GTFB\_seg and GCFB\_seg) Features using Different Classifiers.}
  \label{tab:stm_segmental}
  \small
  \setlength{\tabcolsep}{6pt}
  \renewcommand{\arraystretch}{1.3}
  \begin{tabular}{l c c}
    \hline
 \textbf{Feature Type} & \textbf{Classifier} & \textbf{Accuracy (\%)} \\ 
    \hline
    $\mathrm{STM}_{\mathrm{seg}}$ (GTFB)& SVM & 67.0 \\
     $\mathrm{STM}_{\mathrm{seg}}$ (GCFB) & SVM & 67.0 \\
    \hline
   $\mathrm{STM}_{\mathrm{seg}}$ (GTFB) & KNN & 60.0 \\
     $\mathrm{STM}_{\mathrm{seg}}$ (GCFB) & KNN & 60.0 \\
    \hline
    $\mathrm{STM}_{\mathrm{seg}}$ (GTFB) & Extra Trees & 69.0 \\
     $\mathrm{STM}_{\mathrm{seg}}$ (GCFB) & Extra Trees & \textbf{71.0} \\
    \hline
  \end{tabular}
\end{table}

\begin{table}[htbp]
\centering
\caption{Comparison of Timbral, Mel-spectral, and STM-based Features with Subjective Test Results.}
\label{tab:combined_comparison}
\small
\setlength{\tabcolsep}{6pt}
\renewcommand{\arraystretch}{1.4}
\begin{tabular}{l c c}
\hline
\textbf{Feature Type} & \textbf{Classifier} & \textbf{Accuracy (\%)} \\ 
\hline
Timbral & SVM & 62.0 \\
Timbral & KNN & 58.0 \\
Timbral & Extra Trees & \textbf{65.0} \\
\hline
Mel-Spec & SVM & 51.0 \\
GTFB & SVM & 55.0 \\
GCFB & SVM & 60.0 \\
\hline
STM(GTFB) & KNN & 68.0 \\
STM(GCFB) & KNN & \textbf{69.0} \\
\hline
 $\mathrm{STM}_{\mathrm{seg}}$ (GTFB) & Extra Trees & 69.0 \\
 $\mathrm{STM}_{\mathrm{seg}}$ (GCFB) & Extra Trees & \textbf{71.0} \\
\hline
Subjective Test & Human Evaluation & \textbf{70.0} \\
\hline
\end{tabular}
\end{table}

\subsection{Comparison with Other Methods}
Table~\ref{tab:combined_comparison} compares the overall performance of STM-based representations with other approaches, including timbral features, Mel-spectral features, GTFB, GCFB, and human subjective evaluation. Among all features, the GCFB (60\%) outperformed both GTFB (55\%) and Mel-spectral (51\%), indicating that the gammachirp filterbank’s asymmetric and level-dependent frequency response provides a closer approximation to the human auditory periphery. Timbral features also performed competitively, achieving up to 65\% with Extra Trees, though they primarily represent static spectral attributes such as brightness, roughness, and spectral centroid, and thus lack explicit temporal modulation information.

\begin{figure*}[htbp]
\centering
\begin{minipage}[b]{0.32\textwidth}
  \centering
  \includegraphics[width=\linewidth]{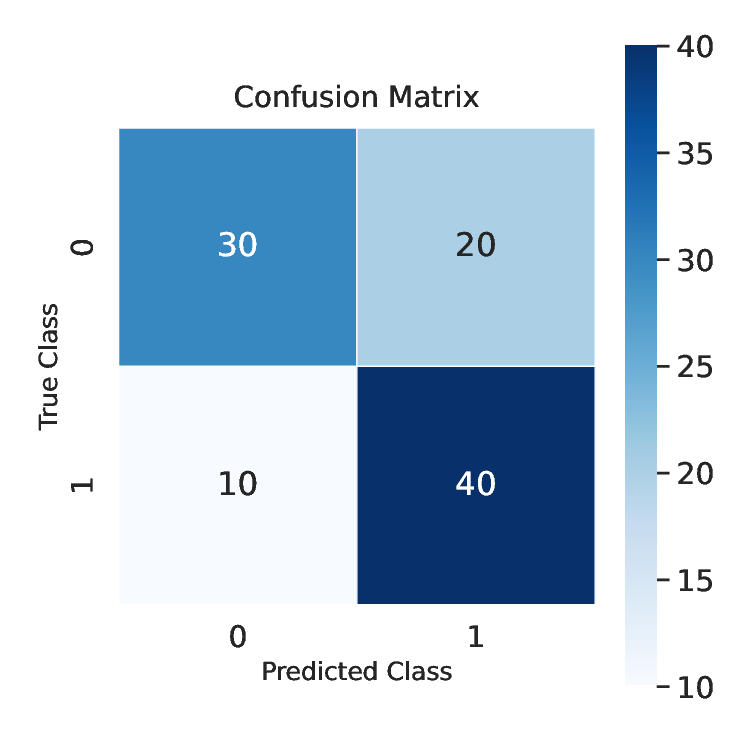}
  \vspace{.5pt}
 {\small (a) Human}
\end{minipage}
\hfill
\begin{minipage}[b]{0.32\textwidth}
  \centering
  \includegraphics[width=\linewidth]{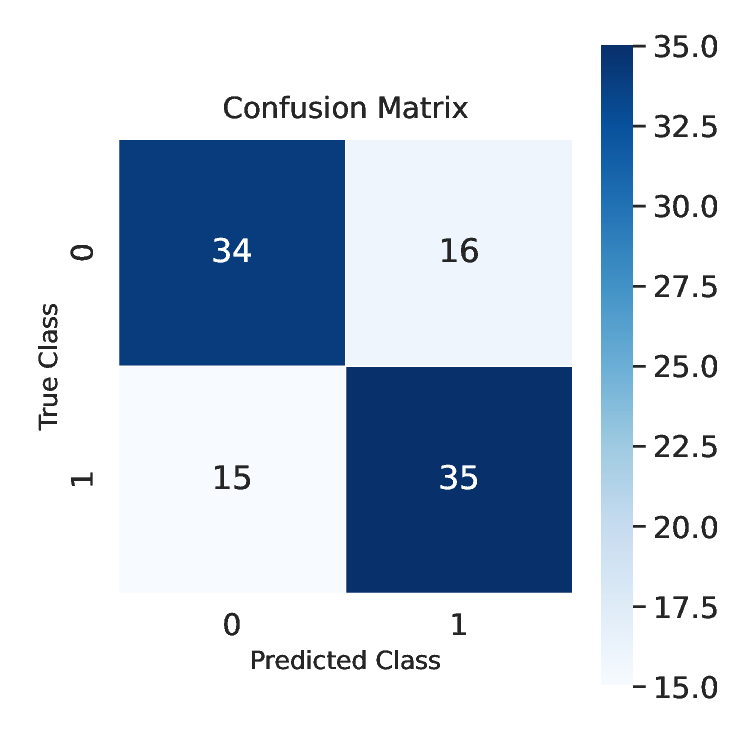}
  \vspace{.5pt}
 {\small (b) $\mathrm{STM}_{\mathrm{seg}}$(GTFB)}
\end{minipage}
\hfill
\begin{minipage}[b]{0.32\textwidth}
  \centering
  \includegraphics[width=\linewidth]{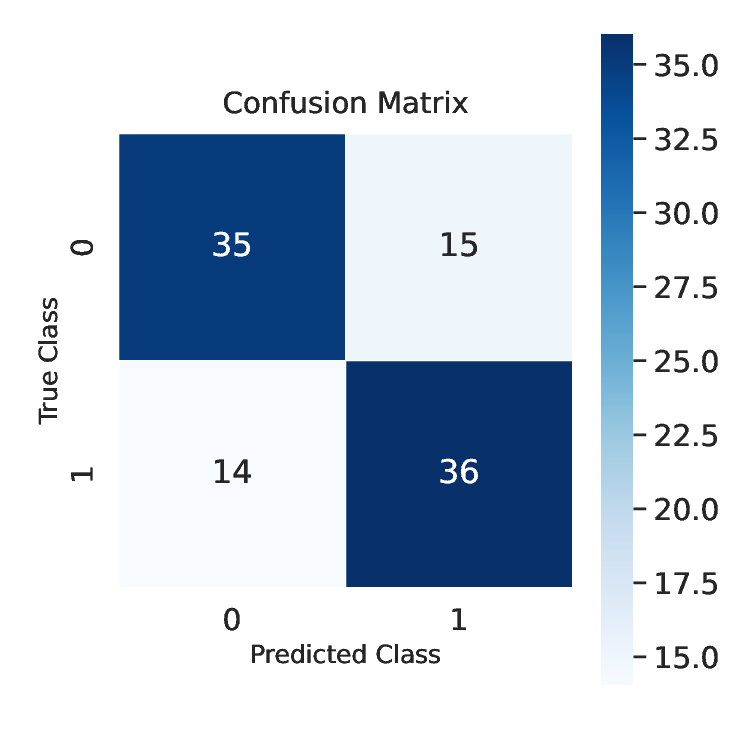}
  \vspace{.5pt}
 {\small (c) $\mathrm{STM}_{\mathrm{seg}}$(GCFB)}
\end{minipage}

\vspace{5pt}
\caption{Confusion matrices showing classification performance for human evaluation and STM-based representations. The STM(GCFB) model demonstrates perceptual alignment with human judgments, particularly in correctly identifying imitated speech.}
\label{mel}
\end{figure*}

In contrast, STM-based features incorporating temporal and spectral modulation yielded significantly higher accuracies. The STM(GCFB) with KNN combination achieved 69\%, nearly matching the human subjective test result (70\%), while the $\mathrm{STM}_{\mathrm{seg}}$(GCFB) with Extra Trees further improved to 71\%. The corresponding confusion matrices in Fig.~\ref{mel} illustrate that STM-based systems produce balanced classification performance across both genuine and imitated speech, closely reflecting human perceptual discrimination patterns observed in subjective evaluation. 

This close alignment between machine and human evaluations underscores the perceptual validity of the proposed auditory-inspired STM framework. 

Unlike conventional acoustic features, STM representations analyze how spectro-temporal modulation patterns dynamically evolve across time and frequency, resembling neural encoding processes in the auditory cortex. 

By modeling temporal and spectral modulations associated with perceptually relevant variations in the speech signal, STM features capture perceptually important spectro–temporal patterns that human listeners rely on when distinguishing genuine speech from imitated speech.

The progression of results from static to dynamic representations emphasizes the importance of modulation-domain processing for perceptually grounded imitation detection. As shown in Table~\ref{tab:combined_comparison} and Fig.~\ref{mel}, the proposed $\mathrm{STM}_{\mathrm{seg}}$(GCFB) representation set achieves a machine performance (71\%) nearly identical to human evaluation (70\%), demonstrating that auditory-inspired and segmentally modulated STM features provide the most perceptually aligned representation for human-imitated speech detection. 

\section{General Discussion}
\label{General Discussion}
The experimental results provide detailed insights of the relationship between human listening tests based on auditory perception and the automatic detection of human-imitated speech using auditory-inspired STM representations, demonstrating strong alignment between human subjective evaluation and the proposed computational framework. In previous work \cite{zaman2025ability}, subjective listening tests showed that human listeners achieved an overall accuracy of approximately 70\% in distinguishing genuine speech from human-imitated speech. This finding indicates that the task depends on perceptually discriminative cues rather than clearly separable acoustic features.

As an expansion of \cite{zaman2025ability}, this study explores the effectiveness of auditory-inspired representations in modeling perceptually relevant characteristics. To this end, conventional acoustic representations, such as Mel-spectral and timbral attributes, are first evaluated as baseline approaches. The results indicate that these representations are limited in capturing the complex characteristics of human-imitated speech. While they primarily describe static spectral properties, they do not adequately represent the dynamic variations that are critical for perceptual discrimination.

In contrast, the proposed STM representations provide a more effective framework by explicitly modeling how spectral content evolves over time. This is particularly important, as human imitation often preserves the overall spectral structure while introducing subtle inconsistencies in temporal and modulation patterns. By capturing both temporal and spectral modulations, STM representations better reflect the underlying mechanisms of human auditory perception.

Furthermore, the incorporation of segmental analysis leads to additional performance improvements. The Segmental-STM approach captures short-term and time-varying modulation patterns that are often averaged out in utterance-level representations. As demonstrated in the results, this approach achieves a classification accuracy of 71\%, slightly surpassing the human performance of 70\%. This finding highlights the importance of short-term modulation cues in distinguishing genuine speech from human imitation.

Another important observation is the strong agreement between model predictions and human subjective evaluation. The confusion matrices reveal similar classification patterns, indicating that the proposed framework captures perceptually relevant characteristics utilized by human listeners. This alignment supports the hypothesis that human auditory perception relies on spectro-temporal modulation cues rather than purely static acoustic information.

Overall, these findings demonstrate that auditory-inspired STM representations provide a perceptually grounded and effective approach for detecting human-imitated speech. By bridging the gap between human listening behavior and computational modeling, this work emphasizes the importance of modulation-domain analysis for tasks involving subtle perceptual differences. The results also suggest that integrating biologically inspired processing principles can lead to more robust and interpretable speech analysis systems.

\section{Conclusion}
\label{Conclusion}
In this work, we proposed STM-based representations and experimentally demonstrated that they provide strong discriminative capability for distinguishing genuine speech from human imitated speech. In particular, STM derived from the GCFB consistently outperformed those based on the GTFB, indicating that level-dependent cochlear modeling contributes to more effective representation of perceptually relevant modulation structures. Furthermore, the proposed segmental-STM representation improved detection performance by capturing short-term temporal variations, highlighting the importance of temporal dynamics in imitation-based speech analysis.

An important observation of this study is the close correspondence between human perceptual accuracy and machine detection performance. Achieving classification results comparable to those of human listeners indicates that the proposed STM-based framework captures key spectro–temporal cues utilized by the human auditory system when evaluating speech authenticity. This alignment supports the perceptual validity of the proposed approach and demonstrates its effectiveness for detecting imitation-based speech forgery, where artifact-oriented or synthesis-specific methods are often limited.

Future work will investigate the integration of additional perceptually motivated features to further enhance robustness and discriminative capability.

\section{Acknowledgment}
This work was partially supported by JSPS KAKENHI (25H01139) and JST Program for co-creating startup ecosystem (JPMJSF2318).

\bibliographystyle{IEEEtran}
\bibliography{refs}

\begin{thebibliography}{10}
\providecommand{\url}[1]{#1}
\csname url@samestyle\endcsname
\providecommand{\newblock}{\relax}
\providecommand{\bibinfo}[2]{#2}
\providecommand{\BIBentrySTDinterwordspacing}{\spaceskip=0pt\relax}
\providecommand{\BIBentryALTinterwordstretchfactor}{4}
\providecommand{\BIBentryALTinterwordspacing}{\spaceskip=\fontdimen2\font plus
\BIBentryALTinterwordstretchfactor\fontdimen3\font minus \fontdimen4\font\relax}
\providecommand{\BIBforeignlanguage}[2]{{%
\expandafter\ifx\csname l@#1\endcsname\relax
\typeout{** WARNING: IEEEtran.bst: No hyphenation pattern has been}%
\typeout{** loaded for the language `#1'. Using the pattern for}%
\typeout{** the default language instead.}%
\else
\language=\csname l@#1\endcsname
\fi
#2}}
\providecommand{\BIBdecl}{\relax}
\BIBdecl

\bibitem{marras2023dictionary}
M.~Marras, P.~Korus, A.~Jain, and N.~Memon, ``Dictionary attacks on speaker verification,'' \emph{IEEE Transactions on Information Forensics and Security}, vol.~18, pp. 773--788, 2023.

\bibitem{wubet2025speaker}
Y.~A. Wubet and K.-Y. Lian, ``Speaker anonymization for voice biometrics protection using voice conversion and multi-target speaker voice fusion,'' \emph{IEEE Transactions on Information Forensics and Security}, 2025.

\bibitem{khan2023securing}
A.~Khan and K.~M. Malik, ``Securing voice biometrics: One-shot learning approach for audio deepfake detection,'' in \emph{2023 IEEE international workshop on information forensics and security (WIFS)}.\hskip 1em plus 0.5em minus 0.4em\relax IEEE, 2023, pp. 1--6.

\bibitem{ballesteros2021deep4snet}
D.~M. Ballesteros, Y.~Rodriguez-Ortega, D.~Renza, and G.~Arce, ``Deep4snet: deep learning for fake speech classification,'' \emph{Expert Systems with Applications}, vol. 184, p. 115465, 2021.

\bibitem{unoki2024deepfake}
{M. Unoki}, {K. Li}, {A. Chaiwongyen}, {Q.-H. Nguyen}, and {K. Zaman}, ``Deepfake speech detection: approaches from acoustic features related to auditory perception to deep neural networks,'' \emph{IEICE Transactions on Information and Systems}, 2024.

\bibitem{zaman2024hybrid}
K.~Zaman, I.~J. Samiul, M.~Sah, C.~Direkoglu, S.~Okada, and M.~Unoki, ``Hybrid transformer architectures with diverse audio features for deepfake speech classification,'' \emph{IEEE Access}, vol.~12, pp. 149\,221--149\,237, 2024.

\bibitem{kanwal2024fake}
T.~Kanwal, R.~Mahum, A.~M. AlSalman, M.~Sharaf, and H.~Hassan, ``Fake speech detection using vggish with attention block,'' \emph{EURASIP Journal on Audio, Speech, and Music Processing}, vol. 2024, no.~1, p.~35, 2024.

\bibitem{wang2017tacotron}
Y.~Wang, R.~Skerry-Ryan, D.~Stanton, Y.~Wu, R.~J. Weiss, N.~Jaitly, Z.~Yang, Y.~Xiao, Z.~Chen, S.~Bengio \emph{et~al.}, ``Tacotron: Towards end-to-end speech synthesis,'' in \emph{Proc. Interspeech 2017}, 2017, pp. 4006--4010.

\bibitem{shen2018natural}
J.~Shen, R.~Pang, R.~J. Weiss, M.~Schuster, N.~Jaitly, Z.~Yang, Z.~Chen, Y.~Zhang, Y.~Wang, R.~Skerrv-Ryan \emph{et~al.}, ``Natural tts synthesis by conditioning wavenet on mel spectrogram predictions,'' in \emph{2018 IEEE international conference on acoustics, speech and signal processing (ICASSP)}.\hskip 1em plus 0.5em minus 0.4em\relax IEEE, 2018, pp. 4779--4783.

\bibitem{van2016wavenet}
A.~van~den Oord, S.~Dieleman, H.~Zen, K.~Simonyan, O.~Vinyals, A.~Graves, N.~Kalchbrenner, A.~Senior, and K.~Kavukcuoglu, ``Wavenet: A generative model for raw audio,'' in \emph{Proc. SSW 2016}, 2016, pp. 125--125.

\bibitem{yamamoto2020parallel}
R.~Yamamoto, E.~Song, and J.-M. Kim, ``Parallel wavegan: A fast waveform generation model based on generative adversarial networks with multi-resolution spectrogram,'' in \emph{ICASSP 2020-2020 IEEE International Conference on Acoustics, Speech and Signal Processing (ICASSP)}.\hskip 1em plus 0.5em minus 0.4em\relax IEEE, 2020, pp. 6199--6203.

\bibitem{kumar2019melgan}
K.~Kumar, R.~Kumar, T.~De~Boissiere, L.~Gestin, W.~Z. Teoh, J.~Sotelo, A.~De~Brebisson, Y.~Bengio, and A.~C. Courville, ``Melgan: Generative adversarial networks for conditional waveform synthesis,'' \emph{Advances in neural information processing systems}, vol.~32, 2019.

\bibitem{kong2020diffwave}
Z.~Kong, W.~Ping, J.~Huang, K.~Zhao, and B.~Catanzaro, ``Diffwave: A versatile diffusion model for audio synthesis,'' \emph{arXiv preprint arXiv:2009.09761}, 2020.

\bibitem{kinnunen2018automatic}
T.~Kinnunen, Z.~Wu, E.~Nicholas~Evans, and J.~Yamagishi, ``Automatic speaker verification spoofing and countermeasures challenge (asvspoof 2015) database,'' 2018.

\bibitem{kinnunen20182nd}
T.~Kinnunen, M.~Sahidullah, E.~H{\'e}ctor~Delgado, E.~Massimiliano~Todisco, E.~Nicholas~Evans, J.~Yamagishi, and K.~A. Lee, ``The 2nd automatic speaker verification spoofing and countermeasures challenge (asvspoof 2017) database, version 2,'' 2018.

\bibitem{kinnunen2017asvspoof}
T.~Kinnunen, M.~Sahidullah, H.~Delgado, M.~Todisco, N.~Evans, J.~Yamagishi, and K.~A. Lee, ``The asvspoof 2017 challenge: Assessing the limits of replay spoofing attack detection,'' in \emph{Interspeech 2017}.\hskip 1em plus 0.5em minus 0.4em\relax International Speech Communication Association, 2017, pp. 2--6.

\bibitem{wang2020asvspoof}
X.~Wang, J.~Yamagishi, M.~Todisco, H.~Delgado, A.~Nautsch, N.~Evans, M.~Sahidullah, V.~Vestman, T.~Kinnunen, K.~A. Lee \emph{et~al.}, ``Asvspoof 2019: A large-scale public database of synthesized, converted and replayed speech,'' \emph{Computer Speech \& Language}, vol.~64, p. 101114, 2020.

\bibitem{liu2023asvspoof}
X.~Liu, X.~Wang, M.~Sahidullah, J.~Patino, H.~Delgado, T.~Kinnunen, M.~Todisco, J.~Yamagishi, N.~Evans, A.~Nautsch \emph{et~al.}, ``Asvspoof 2021: Towards spoofed and deepfake speech detection in the wild,'' \emph{IEEE/ACM Transactions on Audio, Speech, and Language Processing}, vol.~31, pp. 2507--2522, 2023.

\bibitem{wang2024asvspoof}
X.~Wang, H.~Delgado, H.~Tak, J.-w. Jung, H.-j. Shim, M.~Todisco, I.~Kukanov, X.~Liu, M.~Sahidullah, T.~H. Kinnunen \emph{et~al.}, ``Asvspoof 5: crowdsourced speech data, deepfakes, and adversarial attacks at scale,'' in \emph{Proc. ASVspoof 2024}, 2024, pp. 1--8.

\bibitem{yi2022add}
J.~Yi, R.~Fu, J.~Tao, S.~Nie, H.~Ma, C.~Wang, T.~Wang, Z.~Tian, Y.~Bai, C.~Fan \emph{et~al.}, ``Add 2022: the first audio deep synthesis detection challenge,'' in \emph{In Proc. ICASSP 2022-2022 IEEE International Conference on Acoustics, Speech and Signal Processing (ICASSP)}.\hskip 1em plus 0.5em minus 0.4em\relax IEEE, 2022, pp. 9216--9220.

\bibitem{yi2023add}
J.~Yi, J.~Tao, R.~Fu, X.~Yan, C.~Wang, T.~Wang, C.~Y. Zhang, X.~Zhang, Y.~Zhao, Y.~Ren \emph{et~al.}, ``Add 2023: the second audio deepfake detection challenge,'' in \emph{CEUR Workshop Proceedings}, vol. 3597, 2023, pp. 125--130.

\bibitem{reimao2019dataset}
R.~Reimao and V.~Tzerpos, ``For: A dataset for synthetic speech detection,'' in \emph{2019 International Conference on Speech Technology and Human-Computer Dialogue (SpeD)}.\hskip 1em plus 0.5em minus 0.4em\relax IEEE, 2019, pp. 1--10.

\bibitem{ma2024cfad}
H.~Ma, J.~Yi, C.~Wang, X.~Yan, J.~Tao, T.~Wang, S.~Wang, and R.~Fu, ``Cfad: A chinese dataset for fake audio detection,'' \emph{Speech Communication}, vol. 164, p. 103122, 2024.

\bibitem{zaidi2021touch}
A.~Z. Zaidi, C.~Y. Chong, Z.~Jin, R.~Parthiban, and A.~S. Sadiq, ``Touch-based continuous mobile device authentication: State-of-the-art, challenges and opportunities,'' \emph{Journal of Network and Computer Applications}, vol. 191, p. 103162, 2021.

\bibitem{muller2022does}
N.~M{\"u}ller, P.~Czempin, F.~Dieckmann, A.~Froghyar, and K.~B{\"o}ttinger, ``Does audio deepfake detection generalize?'' in \emph{International Speech Communication Association (INTERSPEECH Annual Conference) 2022}, 2022.

\bibitem{yi2021half}
J.~Yi, Y.~Bai, J.~Tao, H.~Ma, Z.~Tian, C.~Wang, T.~Wang, and R.~Fu, ``Half-truth: A partially fake audio detection dataset,'' in \emph{Interspeech}, 2021.

\bibitem{frank2021wavefake}
J.~Frank and L.~Sch{\"o}nherr, ``{WaveFake: A Data Set to Facilitate Audio Deepfake Detection},'' in \emph{Thirty-fifth Conference on Neural Information Processing Systems Datasets and Benchmarks Track}, 2021.

\bibitem{kominek2004cmu}
J.~Kominek and A.~W. Black, ``The cmu arctic speech databases,'' in \emph{Fifth ISCA workshop on speech synthesis}, 2004.

\bibitem{bird2023real}
J.~J. Bird and A.~Lotfi, ``Real-time detection of ai-generated speech for deepfake voice conversion,'' \emph{arXiv preprint arXiv:2308.12734}, 2023.

\bibitem{zhang2022partialspoof}
L.~Zhang, X.~Wang, E.~Cooper, N.~Evans, and J.~Yamagishi, ``The partialspoof database and countermeasures for the detection of short fake speech segments embedded in an utterance,'' \emph{IEEE/ACM Transactions on Audio, Speech, and Language Processing}, vol.~31, pp. 813--825, 2022.

\bibitem{xie2025codecfake}
Y.~Xie, Y.~Lu, R.~Fu, Z.~Wen, Z.~Wang, J.~Tao, X.~Qi, X.~Wang, Y.~Liu, H.~Cheng \emph{et~al.}, ``The codecfake dataset and countermeasures for the universally detection of deepfake audio,'' \emph{IEEE Transactions on Audio, Speech and Language Processing}, 2025.

\bibitem{zhang2021fmfcc}
Z.~Zhang, Y.~Gu, X.~Yi, and X.~Zhao, ``Fmfcc-a: a challenging mandarin dataset for synthetic speech detection,'' in \emph{International Workshop on Digital Watermarking}.\hskip 1em plus 0.5em minus 0.4em\relax Springer, 2021, pp. 117--131.

\bibitem{lorenzo2018voice}
J.~Lorenzo-Trueba, J.~Yamagishi, T.~Toda, D.~Saito, F.~Villavicencio, T.~Kinnunen, and Z.~Ling, ``The voice conversion challenge 2018: Promoting development of parallel and nonparallel methods,'' in \emph{The Speaker and Language Recognition Workshop}.\hskip 1em plus 0.5em minus 0.4em\relax ISCA, 2018, pp. 195--202.

\bibitem{almutairi2022review}
Z.~Almutairi and H.~Elgibreen, ``A review of modern audio deepfake detection methods: challenges and future directions,'' \emph{Algorithms}, vol.~15, no.~5, p. 155, 2022.

\bibitem{kim2018vocal}
B.~Kim, M.~Ghei, B.~Pardo, and Z.~Duan, ``Vocal imitation set: a dataset of vocally imitated sound events using the audioset ontology.'' in \emph{DCASE}, 2018, pp. 148--152.

\bibitem{rodriguez2020machine}
Y.~Rodr{\'\i}guez-Ortega, D.~M. Ballesteros, and D.~Renza, ``A machine learning model to detect fake voice,'' in \emph{International Conference on Applied Informatics}.\hskip 1em plus 0.5em minus 0.4em\relax Springer, 2020, pp. 3--13.

\bibitem{lataifeh2020arabic}
M.~Lataifeh, A.~Elnagar, I.~Shahin, and A.~B. Nassif, ``Arabic audio clips: Identification and discrimination of authentic cantillations from imitations,'' \emph{Neurocomputing}, vol. 418, pp. 162--177, 2020.

\bibitem{mehrabi2019vocal}
A.~Mehrabi, S.~Dixon, and M.~Sandler, ``Vocal imitation of percussion sounds: On the perceptual similarity between imitations and imitated sounds,'' \emph{Plos one}, vol.~14, no.~7, p. e0219955, 2019.

\bibitem{zetterholm2007detection}
E.~Zetterholm, ``Detection of speaker characteristics using voice imitation,'' \emph{Speaker classification II: selected projects}, pp. 192--205, 2007.

\bibitem{hautamaki2015automatic}
R.~G. Hautam{\"a}ki, T.~Kinnunen, V.~Hautam{\"a}ki, and A.-M. Laukkanen, ``Automatic versus human speaker verification: The case of voice mimicry,'' \emph{Speech Communication}, vol.~72, pp. 13--31, 2015.

\bibitem{zaman2025ability}
K.~Zaman, K.~Li, I.~J. Samiul, Y.~Uezu, S.~Kidani, and M.~Unoki, ``Ability of human auditory perception to distinguish human-imitated speech,'' \emph{IEEE Access}, 2025.

\bibitem{flinker2019spectrotemporal}
A.~Flinker, W.~K. Doyle, A.~D. Mehta, O.~Devinsky, and D.~Poeppel, ``Spectrotemporal modulation provides a unifying framework for auditory cortical asymmetries,'' \emph{Nature human behaviour}, vol.~3, no.~4, pp. 393--405, 2019.

\bibitem{wu2013synthetic}
Z.~Wu, X.~Xiao, E.~S. Chng, and H.~Li, ``Synthetic speech detection using temporal modulation feature,'' in \emph{2013 IEEE International Conference on Acoustics, Speech and Signal Processing}.\hskip 1em plus 0.5em minus 0.4em\relax IEEE, 2013, pp. 7234--7238.

\bibitem{cheng2023analysis}
H.~Cheng, C.~O. Mawalim, K.~Li, L.~Wang, and M.~Unoki, ``Analysis of spectro-temporal modulation representation for deep-fake speech detection,'' in \emph{2023 Asia Pacific Signal and Information Processing Association Annual Summit and Conference (APSIPA ASC)}.\hskip 1em plus 0.5em minus 0.4em\relax IEEE, 2023, pp. 1822--1829.

\bibitem{li2025machine}
K.~Li, K.~Zaman, X.~Li, M.~Akagi, J.~Dang, and M.~Unoki, ``Machine anomalous sound detection using spectral-temporal modulation representations derived from machine-specific filterbanks,'' \emph{IEEE Transactions on Audio, Speech and Language Processing}, 2025.

\bibitem{bance2007hearing}
M.~Bance, ``Hearing and aging.'' \emph{CMAJ: Canadian Medical Association Journal}, vol. 176, no.~7, pp. 925--928, 2007.

\bibitem{chi2012robust}
T.-S. Chi, L.-Y. Yeh, and C.-C. Hsu, ``Robust emotion recognition by spectro-temporal modulation statistic features,'' \emph{Journal of Ambient Intelligence and Humanized Computing}, vol.~3, no.~1, pp. 47--60, 2012.

\bibitem{patterson1988efficient}
R.~D. Patterson, I.~Nimmo-Smith, J.~Holdsworth, and P.~Rice, ``An efficient auditory filterbank based on the gammatone function,'' Applied Psychology Unit, Cambridge, UK, APU Report 2341, 1988.

\bibitem{irino1997gammachirp}
T.~Irino and R.~Patterson, ``A time-domain, level-dependent auditory filter: The gammachirp,'' \emph{The Journal of the Acoustical Society of America}, vol. 101, no.~1, pp. 412--419, 1997.

\bibitem{irino1999analysis}
T.~Irino and M.~Unoki, ``An analysis/synthesis auditory filterbank based on an iir implementation of the gammachirp,'' \emph{Journal of the Acoustical Society of Japan (E)}, vol.~20, no.~6, pp. 397--406, 1999.

\end{thebibliography}

\vfill

\end{document}